# Modeling and forecasting the COVID-19 temporal spread in Greece: an exploratory approach based on complex network defined splines.

Konstantinos Demertzis[1]*, Dimitrios Tsiotas[1,2,3], and Lykourgos Magafas[1]

[1] Laboratory of Complex Systems, Department of Physics, Faculty of Sciences, International Hellenic University, Kavala Campus, St. Loukas, 65404, Greece; kdemertzis@teiemt.gr; magafas@teikav.edu.gr;

[2] Department of Regional and Economic Development, Agricultural University of Athens, Greece, Nea Poli, Amfissa, 33100, Greece; tsiotas@aua.gr;

[3] Department of Planning and Regional Development, University of Thessaly, Pedion Areos, Volos, 38334, Greece; tsiotas@uth.gr;

* Correspondence: kdemertzis@teiemt.gr;

**Abstract:** Within the complex framework of anti-COVID-19 health management, where the criteria of diagnostic testing, the availability of public-health resources and services, and the applied anti-COVID-19 policies vary between countries, the reliability and the accuracy in the modeling of temporal spread can be proven effective in the worldwide fight against the disease. This paper applies an exploratory time-series analysis to the evolution of the disease in Greece, which currently suggests a success story of COVID-19 management. The proposed method builds on a recent conceptualization of detecting connective communities in a time-series and develops a novel spline regression model where the knot vector is determined by the community detection in the complex network. Overall, the study contributes to the COVID-19 research by proposing a free of disconnected past-data and reliable framework of forecasting, which can facilitate decision-making and management of the available health resources.

## 1. INTRODUCTION

The coronavirus disease 2019, abbreviated as COVID-19, is a contagious disease caused by the severe acute respiratory syndrome coronavirus 2 (SARS-CoV-2), which frequently causes fever, cough, and dyspnea, and less frequently muscle pains and neck-related problems (Ahmed et al., 2020; Lescure et al., 2020; Xu et al., 2020b). The virus is mainly transmitted to humans through respiratory channels and the majority of patients is asymptomatic or has soft symptoms to the disease, but certain cases develop either pneumonia, the worst aspect of which is the fatal Acute Respiratory Distress Syndrome (ARDS), or multi-organic deficiency (Xu et al., 2020b). The time from exposition to the appearance of symptoms ranges from 2 to 14 days, with a 5-day average, the long range of which is affected by the relevance of the disease, in its asymptomatic or soft symptomatology aspect, with the common cold (Fang et al., 2020; Heymann and Shindo, 2020). An amount of 25% of - 30% of patients worsens after the 14$^{th}$ day of exposition, showing respiratory infection, whereas an amount of 83% of patients develops lymphopenia (Tan et al., 2020). The disease is also observed in children, by usually with soft symptoms (Liu et al., 2020; Qiu et al., 2020).

The COVID-19 is detected either by laboratory methods, usually by the method of Polymerase Chain Reaction (PCR) (Bai et al., 2020), where the sample is received from the rhino-laryngology region, or just by clinical methods, by evaluating combinations of symptoms (at least of two major), danger-factors, and indications of the chest radiogram, in conjunction with the history of the patients' contacts and movements (Bai et al., 2020; Ye et al., 2020). Currently, since no vaccine or cure for the disease is available (Lurie et al., 2020), the major efforts of the medical community are focusing on the management of symptoms, while of the governance are focusing on the management of public-health resources and on the prevention management of the disease. Within the context that COVID-19 is a particularly airborne contagious disease, medical directives to the public highlight the need for careful personal body hygiene, while government policies (Tsiotas and Magafas, 2020) impose severe restrictions of mobility, gathering, transportation, and trade activities.

In particular, starting to the mid of December 2019, when COVID-19 emerged in the city of Wuhan, China, up to the 19$^{th}$ of April 2020, the disease was spread up to 210 countries, causing 2,408,123 infections and 165,105 deaths (Roser and Ritchie, 2020). Despite that its spatiotemporal pattern specializes amongst countries worldwide, it is a common feature that the pandemic shows scaling trends worldwide (with couple exemptions the cases of South Korea and China) without illustrating a tendency of stabilization. For instance, up to the 19$^{th}$ of April 2020, North America recorded 43,369 deaths and 820,749 infections, South America 3,850 deaths and 82,310 infections, Asia 14,801 deaths and 383,542 infections, Africa 1,128 deaths and 22,992 infections, and Oceania 83 deaths and 8,150 infections (Our World in Data, 2020; Worldometers, 2020). On the other hand, Europe is more hardly affected by the pandemic because, although it counted the half infections (1,089,659), it recorded over the 60% of the worldwide deaths (101,859), from which an amount of 78.4% (72,196 deaths) were in Italy (23,660), Spain (20,453), France (19,718), and United Kingdom (16,060) (Our World in Data, 2020; Worldometers, 2020).

On the contrary to the worldwide and European status of COVID-19, Greece has a proportion of 240 confirmed infections per million of population, which is almost 35% lower than the global average, which is about 370 infections per million, and 85% than the European average, which is about 1330 infections per million (Our World in Data, 2020; Worldometers, 2020). Within the context of promising a success story in the fight against the disease, this paper develops a novel non-parametric method for the modeling of the evolution of the Greek COVID-19 infection-curve that can facilitate to more accurate forecasting. The proposed method builds on a recent conceptualization of detecting communities of connectivity in a time-series (Tsiotas and Magafas, 2020) and develops a novel model based on the Regression Splines algorithm that is more accurate and reliable in forecasting. The overall approach provides insights of good policy and decision-making practices and management that can facilitate decision-making and management of the available health resources in the fight against COVID-19.

The remainder of the paper is organized as follows; Section 2 reviews the literature in the current analysis of COVID-19 temporal spread, Section 3 applies a descriptive analysis of COVID-19 in Greece, Section 4 describes the methodological framework of the proposed method, Section 5 shows the results of the analysis and discusses them within the context of public-health management and practice, and, finally, in Section 6 conclusions are given.

## 2. LITERATURE REVIEW

The work of Xu et al (2020) is a detailed presentation of COVID-19 records that are extracted from national, regional, municipal health-reports, and web information aiming to contribute to the decision-making for the public health with insightful primary information. This work focuses more on the recording than on the analysis of cases and therefore it exclusively contributes to COVID-19 research as an archive of statistical data. The work of Sarkodie and Owusu (2020) is an insightful time-series analysis examining the interconnection between deaths and infected cases, based on four health indicators of COVID-19, in China. The analysis uses cross-sectional dependence, endogeneity, and unobserved heterogeneity estimation methods, and detects a linear relationship between COVID-19 attributable deaths and confirmed cases whereas a non-linear relationship ruling the nexus between recovery and confirmed cases. This work contributes to the literature with an interesting case-study that is by default restricted to the case of China, to the limited number of indicators used in the analysis, and to the limited time-series dataset that does not facilitate reliable forecasting. Next, the work of Anastassopoulou et al. (2020) proposes a heuristic method for estimating basic epidemiologic parameters, for modeling, and forecasting the COVID-19 spread based on available epidemiologic data. Their approach suggests a reverse forecasting process that builds on spreading scenarios, which reproduce the confirmed cases, and it develops a directed tendency which cannot promise a reliable basis for forecasting. Also, Petropoulos and Makridakis (2020) study the temporal-spread of the disease based on exponential smoothing modeling. Although interesting, this approach is restricted to the insufficient number of past data, on which their exponential model and promises to forecast the tendency of the COVID-19 future spreading model of illness of past, while the fitted curve is calibrated and smoothened in accordance with the foregoing available cases of other countries. Moreover, Fong et al. (2020) presented an interesting forecasting model, based on a polynomial neural network with corrective feedback, which is capable to forecast with satisfactorily accuracy, even in cases of insufficient data availability. Although interesting, this approach should be further tested and compared with alternative established algorithms of similar good accuracy by taking into consideration more than the accuracy criterion for the comparison.

On the other hand, due to the diversity that the phenomenon has in different countries, many researchers were focused on national case-studies of COVID-19 instead of the global case. For instance, Mahase (2020) demonstrated the changes in statistical data of the United Kingdom after the application of the anti-COVID-19 social distance policies. In Italy, many studies were conducted for the modeling of the pandemic (Giuliani et al., 2020; Livingston and Bucher, 2020; Remuzzi and Remuzzi, 2020), due to fatal outbreak that the disease had in the country, which attracted the global attention. The work of Giuliani et al. (2020) is a characteristic early study of modeling the spatiotemporal spread of COVID-19 in Italy, where the analysis builds on statistical modeling but without testing the statistical significance of the research hypothesis. Within the context of epidemiologic research, this incompleteness restricts the contribution of this interesting approach, provided that in epidemiologic studies the goal is to develop an occurrence function (as a measure of association) quantifying a cause-effect relation between a determinant (cause) and its result (effect) and therefore the major concern is to test whether this cause-effect relation is statistically significant.

Greece is an example of timely response in the application of anti-COVID-19 policies that are currently have been proven effective in keeping the infected cases and deaths at relatively low levels (Roser and Ritchie, 2020; Xu et al., 2020; Tsiotas and Magafas, 2020). In particular, the first infection emerged in the country on February 26$^{th}$, 2020, and just three days later the state began applying several policies for the control of the disease (Tsiotas and Magafas, 2020). This timely response have led Greece to be currently considered as a successful case in anti-COVID-19 management, comparatively both to the European and the global cases (Roser and Ritchie, 2020; Xu et al., 2020). At the time that Greece started attracting the global attention, Tsiotas and Magafas (2020) proposed a novel complex network analysis of time-series, based on the visibility algorithm (Lacasa et al., 2008; Tsiotas and Charakopoulos, 2020), for the study of the Greek COVID-19 infection curve. The authors showed that the evolution of the disease in Greece went through five stages of declining dynamics, where saturation trends (represented by a logarithmic pattern) emerged after the 33$^{rd}$ day (29/04/2020). Within the context that Greece promises a success story and an insightful case study, both in epidemiologic and in anti-COVID-19 policy terms, this paper builds on the very recent work of Tsiotas and Magafas (2020) and advances the time-series modeling and forecasting by developing a model based on the Regression Splines algorithm that is more capable to provide accurate predictions of future trends.

## 3. DESCRIPTIVE ANALYSIS OF COVID-19 IN GREECE

The disease of COVID-19 emerged in Greece on 26 February 2020, almost two months after its global emergence (Tsiotas and Magafas, 2020). As it can be observed in Fig.1, within the 54 first days of the pandemic (until 19 April), Greece recorded 2,235 confirmed infected cases (Worldometers, 2020), from which an amount of 56% are men, 25.5% (570 cases) are related with traveling abroad, 42.2% (943 cases) are linked with other confirmed cases, whereas the others are untracked and are still undergoing investigation (NPHOG, 2020). The average age of cases is 49 years (ranging from 1 day until 102 years old), whereas the medium death age is 74 years (ranging from 39 to 95 years) (NPHOG, 2020).

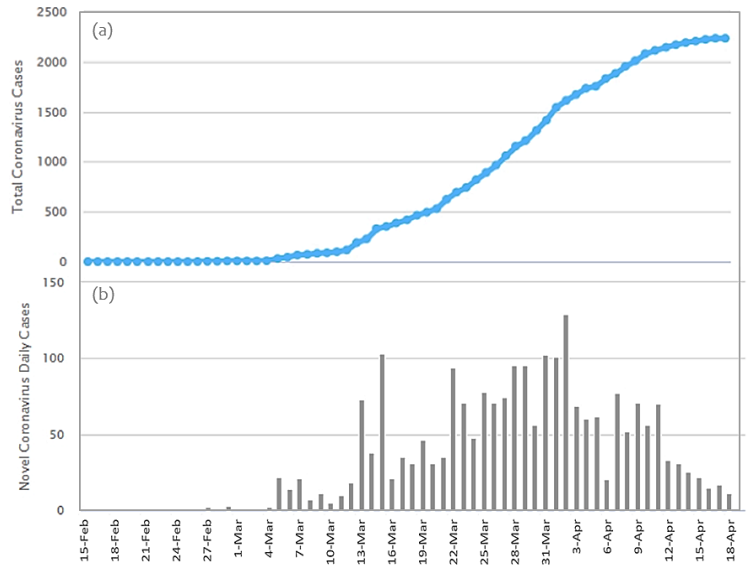

**Fig.1.** The time-series of the COVID-19 infection curve in Greece, for the period 15 Feb 20 - 18 Apr 20. The first infection emerged on 26 Feb 20 (and recorded on 27 Feb 20) (data source: Worldometers, 2020).

In numeric terms, Greece is worldwide at the 58[th] place in number of infected cases and at the 46[th] place in number of deaths (Worldometers, 2020), while, in Europe, is at the 25[th] place in number of cases and at the 22[th] place in deaths. In per capita terms, Greece has currently 13 deaths per million of population (Worldometers, 2020), while the global average is above 16 deaths per million of residents (Our World in Data, 2020). Also, with 67 patients being under a serious-critical situation, Greece is worldwide at the 37[th] place and at the 20[th] place in Europe (NPHOG, 2020; Our World in Data, 2020; Worldometers, 2020). In terms of testing, Greece has conducted 50,771 tests and takes the 56[th] place worldwide and the 23[rd] place in Europe (NPHOG, 2020; Our World in Data, 2020; Worldometers, 2020).

The geographic distribution of the confirmed infected cases in Greece are shown in the map of Fig.2 (NPHOG, 2020), where it can be observed that the majority of infections are concentrated in the metropolitan prefectures of Attiki (6) and Thessaloniki (47), along a vertical axis configured by the prefectures of Kastorias (24), Kozanis (30), and Larissas (33), at central Greece, in the prefecture of Euvoias (12), in the prefectures of Xanthis (50) and Evrou (13), at the north-east country, and at the prefectures of Achaias (1), Heleias (19), and Zakenthou (51), at the south-west Greece, the majority of which are transportation (road, maritime, and air transport) centers.

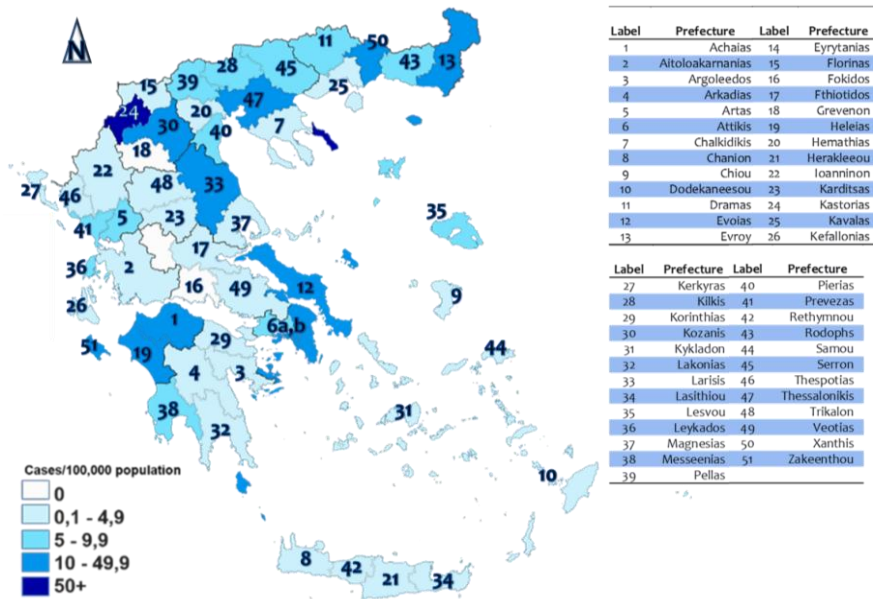

**Fig.2.** Infected cases per million in Greece (source: NPHOG, 2020)

Also, Greece is worldwide in the 68[th] place and in the 26[th] place in Europe regarding the number of patients recovered from COVID-19 (NPHOG, 2020; Our World in Data, 2020; Worldometers, 2020). The daily and cumulative infections of COVID-19 in Greece are shown in Fig.3.

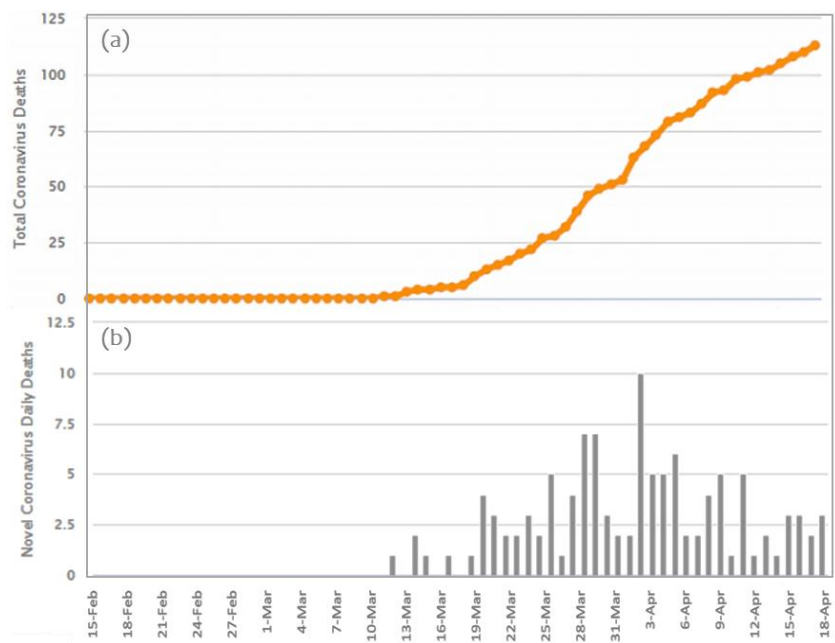

**Fig.3.** The time-series of the COVID-19 death curve in Greece, for the period 15 Feb 20 - 18 Apr 20. The first death was recorded on 12 Mar 20 (data source: Worldometers, 2020).

Next, Fig.4 shows the evolution of the confirmed infected cases in Greece comparatively to the respective recorded deaths. The vertical axis of the diagram is graded at the logarithmic scale, where linear segments illustrate exponential growth of the disease (the slope of the linear growth is proportional to the size of the exponent). As it can be observed, the almost constant offset between the two curves implies that the number of infections and the number of deaths of COVID-19 in Greece are correlated. This interprets that these two indicators follow a similar growth pattern, which is in line with the shape of the curves shown in Figures 1 and 3. Also, a particularly promising observation is the declining growth rates shown in Fig.4 for both the infection and death curves. However, these observations will be statistical tested at a following part of the analysis.

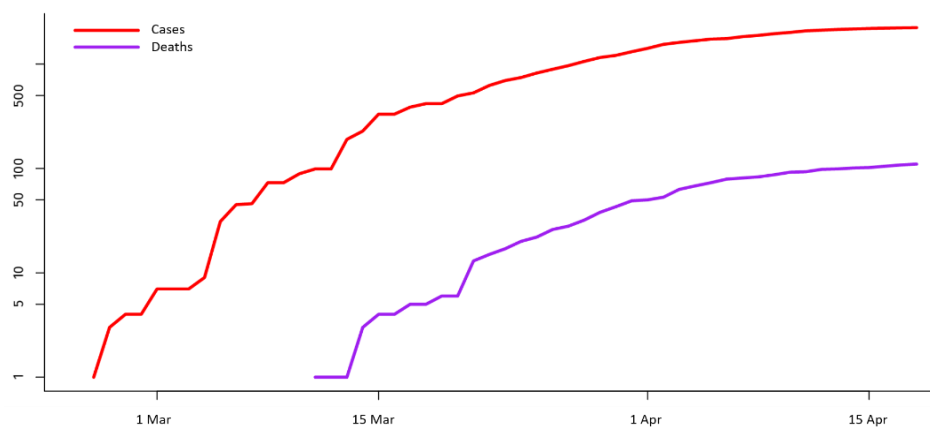

**Fig.4.** Comparative diagram with the time-series of the COVID-19 infection cases versus the recorded deaths in Greece (data source: NPHOG, 2020).

Next, Fig.5 is an aggregate diagram showing the evolution of COVID-19 confirmed (total) cases, new infections, deaths, and recovered on official data extracted from the Greek Ministry of Health (NPHOG, 2020).

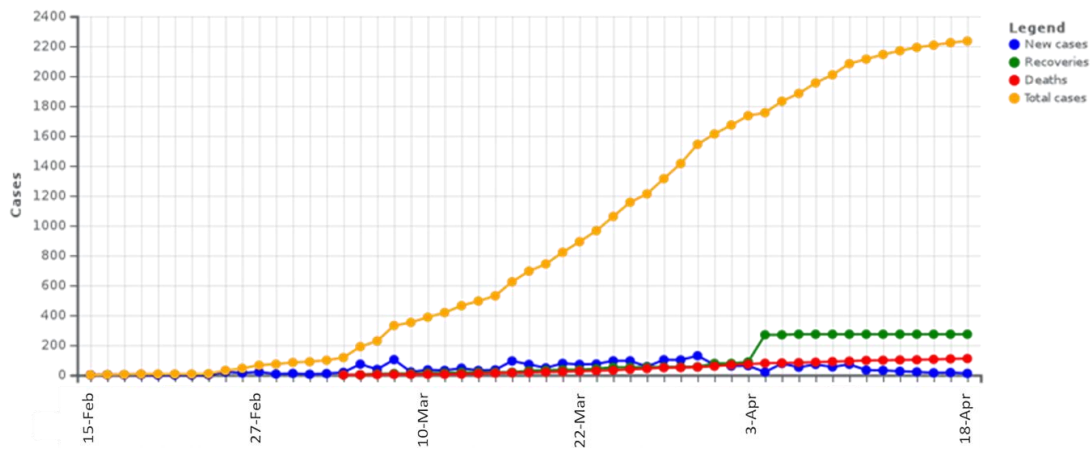

**Fig.5.** The aggregate time-series of the COVID-19 in Greece, showing the number of **confirmed (total) cases**, **new infections**, **deaths**, and **recoveries**, for the period 15 Feb 20 - 18 Apr 20 (data source: Worldometers, 2020).

Within the context of the global outbreak of the disease, the previous descriptive analysis illustrates that Greece suggests a good example for its COVID-19-related sizes, which keep the country at the last places both in the European and the global ranking. However, this good performance has been the result of the timely response and application of anti-COVID-19 policies in Greece (Tsiotas and Magafas, 2020), including a set of measuring of social distancing to prevent the spreading of the disease, In particular, the first anti-COVID-19 policies in Greece were applied after the confirmation of the first three infected cases, which were dated on the 27th of February 2020. On that day, all carnival events were cancelled to prevent an outbreak of the disease. On the 10th of March, the number of total cases reached 89 (MOHG, 2020), the tracking of which revealed that they were mainly related either with travellers originating from Italy or with pilgrims returned from religious excursion to Israel of with travelers from Egypt, along with their contacts (MOHG, 2020). On that day, the government announced to the population directives related to personal hygiene, social distancing, and prevention, while the anti-COVID-19 measures up to then were optional and applicable at the local level (and particularly at the regions with infected cases, such as Heleias-19, Achaias-1 and Zakynthou-51), mainly concerning local suspension of schools, school excursions abroad, and cultural events (Tsiotas and Magafas, 2020). However, on 10th of March, due to the spreading of the disease to multiple regions and due to the disobedience of the citizens to conform with the measures, the government applied more active measures and proceeded to the national suspension of all educational structures (at all ranks), and couple of days later, on the 12th and 13th of March, it proceeded to the suspension of cafeterias, bars, museums, malls and trade centers, sport activities, restaurants, and relevant (MOHG, 2020). On 16th of March, all commercial shops were suspended at the national level, two villages in the regions of Kozani (30) were put into quarantine, and all doctrine and religious activities were suspended (MOHG, 2020). The only active businesses and firms exempted from these measures were primary need suppliers, such as bakeries, supermarkets, pharmacies, and health private services (MOHG, 2020). Aiming to support the anti-COVID-19 policy of social distancing, the government announced, on 18th and 19th of March, a set of measures of 10 billion Euros (€) budgest, which either concerned taxation benefits or regulations or subsidies, for the support of the economy, of companies, and of workers offended by the suspensions and by the social distancing (MOHG, 2020). On the 23rd of March 2020 the government announced national restrictions in transportation, with exemptions the commuting to work, movements for supplies of food, medicines, medical services, and health-gymnastics. However, these exemptions should be documented by identification papers, such as the ID-card or passport and an affirmation paper stating the purpose of movement (MOHG, 2020). At their circulation, provided that he is related with somebody of the acceptable exceptions, the citizens will be supposed to bring their police identity or the passport, as well as some type certification of proportionally aim of locomotion. This measure was applicable until 27th of April 2020 and it was extended until 4th of May 2020 (MOHG, 2020).

In the fight against the disease, the development of more accurate and reliable models in terms of description and prediction can help policy makers to better conceptualize the pandemic and apply proper and more effective policies. Towards this direction, this paper proposes a novel complex-network-based approach of the splines algorithm, which facilitates better epidemiologic modeling and forecasting.

## 4. METHODOLOGY AND DATA

The available data were extracted from the National Public Health Organization of Greece (NPHOG, 2020) and the Ministry of Health of Greece (MOHG, 2020). The variables participating to the analysis include daily cases of the period 26 Feb 2020 until 16 Apr 2020 and are: the day since the first infection in Greece (variable $X_1$: Day), the COVID-19 cumulative infected cases (variable $X_2$: Infections) and cumulative deaths (var. $X_3$: Deaths), the daily infections (var. $X_4$: Daily Infections), the daily deaths (var. $X_5$: Daily Deaths), daily recoveries (var. $X_6$: Daily Recovered), the daily new patients in the Intensive Care Unit (var. $X_7$: ICU), and the daily number of tests (var. $X_8$: Tests). All available variables are shown in table A1 of the Appendix. Each variables is a time-series $x(n)=\{x(1), x(2), \ldots, x(n)\}$ where each node $i=1,2,\ldots,n$ refers to a day since the first infection.

Overall, the analysis examines the dynamics of the Greek COVID-19 infection curve, as it is expressed by the available time-series variable $X_2:X_8$. The study is implemented through a double perspective; the first examines the structural dynamics of one variable ($X_i$, $i=1,\ldots,n$) in comparison with the other available variables $X_j$ (analysis between variables, $i \neq j=1,\ldots,n$), whereas the second examines the time-series pattern configured for a variable $X_i$ (analysis within variable $X_i$, $i=1,\ldots,n$). Towards the first direction, a Pearson's bivariate correlation analysis is applied to the set of the available variables ($X_2: X_8$) and the results are shown in Table 1. As it can be observed, the number of infections ($X_2$) is significantly correlated with all variables except $X_6$ (daily recovered) and the daily number of infections ($X_4$) is significantly correlated with the daily number of deaths ($X_5$) and the patients in ICU ($X_7$). These significant results imply, on the one hand, that the coevolution of the COVID-19 infection curve with variables $X_3:X_5$, $X_7,X_8$ is less than 5% likely to be a matter of chance, and, on the other hand, that the coevolution of the daily COVID-19 infections with the daily number of deaths and the patients in ICU is less than 1% likely to be a matter of chance. In general, the correlation analysis indicate that the evolution of the COVID-19 infections in Greece is very likely to submitted to causality and less likely to be a matter of chance.

Table 1

Results of the Pearson's bivariate correlation analysis (r)

| | | $X_3$ | $X_4$ | $X_5$ | $X_6$ | $X_7$ | $X_8$ |
|---|---|---|---|---|---|---|---|
| Infections ($X_2$) | r(x,y) | 0.979** | 0.317* | 0.612** | 0.138 | -0.316* | 0.585** |
| | Sig. (2-tailed) | 0.000 | 0.020 | 0.000 | 0.321 | 0.020 | 0.000 |
| Deaths ($X_3$) | r(x,y) | 1 | 0.153 | 0.496** | 0.123 | -0.421** | 0.552** |
| | Sig. (2-tailed) | | 0.268 | 0.000 | 0.376 | 0.002 | 0.000 |
| Daily Infections ($X_4$) | r(x,y) | | 1 | 0.487** | -0.038 | 0.358** | 0.232 |
| | Sig. (2-tailed) | | | 0.000 | 0.785 | 0.008 | 0.091 |
| Daily Deaths ($X_5$) | r(x,y) | | | 1 | 0.277* | 0.010 | 0.339* |
| | Sig. (2-tailed) | | | | 0.042 | 0.941 | 0.012 |
| Recovered ($X_6$) | r(x,y) | | | | 1 | -0.133 | 0.003 |
| | Sig. (2-tailed) | | | | | 0.339 | 0.982 |
| ICU ($X_7$) | r(x,y) | | | | | 1 | -0.077 |
| | Sig. (2-tailed) | | | | | | 0.582 |

*. Coefficient significant at the 0.05 level

**. Coefficient significant at the 0.01 level

As far as correlations of other of variables is concerned, Table 1 shows that the number of recoveries ($X_6$) is significantly (but not highly) and positively correlated with the number of daily deaths ($X_5$), expressing a tendency of the Greek health system to get more recoveries when the number of deaths increases. This correlation illustrates and analogy between deaths and recoveries suggesting a variable for further research. Also, the number of patients in ICU ($X_7$) appears significantly and negatively correlated with the number of infections ($X_2$) and deaths ($X_3$), implying that the number of patients in ICU tend to decrease when the number of infections and deaths increases. This observation is rationale since cases of death are removed from the ICU. On the other hand, the number of patients in ICU ($X_7$) is significantly and positively correlated with the number of daily infections ($X_4$), implying that the number of patients entering in ICU tend to increase when the number of daily infections gets bigger. Next, an interesting observation regards the correlations between the number of tests ($X_4$) and variables $X_2$ and $X_3$. Although these correlations $r(X_4,X_2)$ and $r(X_4,X_3)$ are significant and positive, as expected (implying that the number of tests appears proportional to the number of infections and deaths), the numerical values of these coefficients do not appear considerably high (since $r(X_4,X_2)$, $r(X_4,X_2) < 0.6$). Provided that a perfect positive linearity between the number of tests and infections (or deaths) implies a increasing awareness of the health system proportionally to the spread of the disease, the considerable high distance (> 40%) of the correlation coefficients $r(X_4,X_2)$, $r(X_4,X_2)$ from the perfect (positive) linearity can be seen as an aspect of testing ineffectiveness of the health system in Greece. Overall, the correlation analysis showed that different aspects of the disease in Greece are ruled by non-randomness and therefore it provided indications that the evolution of the Greek COVID-19 system is driven by sort-term linear trends. Therefore, a stochastic analysis is further applied for the improvement of the overall systems' determination and thus for the better conceptualization of the dynamics ruling the evolution of COVID-19 in Greece.

*4.1. Regression Analysis*

The first approach for modeling the evolution of the GOVID-19 infection curve builds on regression analysis and generally on the curve fitting approach (Walpole et al., 2012), according to which a parametric curve is fitted to the data of variable $X_2=f(t)$ that bests describes its variability through time. The available types of fitting curves examined in the regression analysis are the linear, quadratic (2$^{nd}$ order polynomial), cubic (3$^{nd}$ order polynomial), power, and logarithmic. All the available types of fitting curves can be generally described by the general multivariate linear regression model expressed by the formula (Walpole et al., 2012):

$$\hat{y} = b_1 x_1 + b_2 x_2 + \cdots + b_n x_n + c = \sum b_i x_i + c \qquad (1),$$

by considering that each independent variable $x_i$ can represent a function of $x$, namely $x_i = f(x)$, as it is shown in relation (2).

$$\hat{y} = \sum b_i f(x_i) + c \qquad (2),$$

The function $f(x)$ can be either logarithmic $f(x)=(\log(x))^m$, or polynomial $f(x)=x^m$, or exponential $f(x)=(\exp\{x\})^m$, or any other. Within this context, the purpose of the regression analysis is estimating the parameters $b_i$ of model (2) that best fit to the observed data $y$, so that to minimize the square differences $y_i - \hat{y}_i$ (Walpole et al., 2012), namely:

$$min\{e = \sum_{i=1}^{n}[y_i - \hat{y}_i]^2 = \sum_{i=1}^{n}[y_i - (\sum b_i f(x_i) + c)]^2\} \qquad (3).$$

The algorithm estimates the beta coefficients ($b_i$) by using the Least-Squares Linear Regression (LSLR) method (Walpole et al., 2012), based on the assumption that the differences $e$ in relation (3) follow the normal distribution $N(0,\sigma_e^2)$. In this paper, the time (days since the first infection, variable $X_1$) are set as the independent variable and each other available variable are set as response variables to the models. In all cases, the simplest form of regression models that best fits the data is chosen. The simplicity criterion regards both the number of the used terms $b_i f(x_i)$ and the polynomial degree $m$. That is, the model with the least possible terms and the lowest possible degree $m < n$-2 (where $n$ is the number of observations in the data-set) is chosen if it best fits the data. The determination ability of each model is expressed by the coefficient of determination $R^2$, which is given by the formula (Norusis, 2008; Walpole et al., 2012):

$$R^2 = 1 - \frac{\sum_{i=1}^{n}(Y_i - \hat{Y}_i)^2}{\sum_{i=1}^{n}(Y_i - \bar{Y}_i)^2} \qquad (4),$$

where $Y_i$ are the observed values of the response (independent) variable, $\hat{Y}_i$ are the estimated values of the response variable, $\bar{Y}$ is the average of the observed values of the response variable and *n* is the number of observations (the length of the variables). The coefficient of determination expresses the amount of the variability of the response variable is expressed by the model $\hat{Y}$ and ranges within the interval [0,1], showing perfect determination when equals to one (Norusis, 2008; Walpole et al., 2012).

Another measure of fitting ability is the root mean square deviation or error (RMSD or RMSE), which calculates the square root of the expected differences between the predicted ($\hat{y}$) and the observed (*y*) values of the model, according to the formula (Norusis, 2008; Walpole et al., 2012):

$$RMSE = \sqrt{MSE(\hat{y})} = \sqrt{E((\hat{y}-y)^2)} \qquad (5),$$

where $E(\cdot)$ is the function of the expected value. The RMSE represents the square root of the second sample moment of the regression residuals (Norusis, 2008; Walpole et al., 2012).

A final measure of fitting ability used in the analysis is the relative absolute error (RAE), which calculates the relative value of the RMSE in accordance with the expected observed values, as it is shown at the formula (Norusis, 2008; Walpole et al., 2012):

$$RMSE = \sqrt{E((\hat{y}-y)^2)}/\sqrt{E((y)^2)} \qquad (6),$$

where $E(\cdot)$ is the function of the expected value. The RAE is often used in machine learning, data mining, and operations management applications, and represents the analogy of the RMSE relatively to the expected value of the observed values.

Within this context, Fig.6 shows the results of the regression analysis applied to the (cumulative) number of infections (dependent variable: $X_2$). As it can be observed, the 3rd order polynomial (cubic) fitting curve best describes the data of the Greek COVID-19 cumulative infections. The last (very recent) part of the cubic curve appears convex implying that the number of cumulative infections tends to saturate.

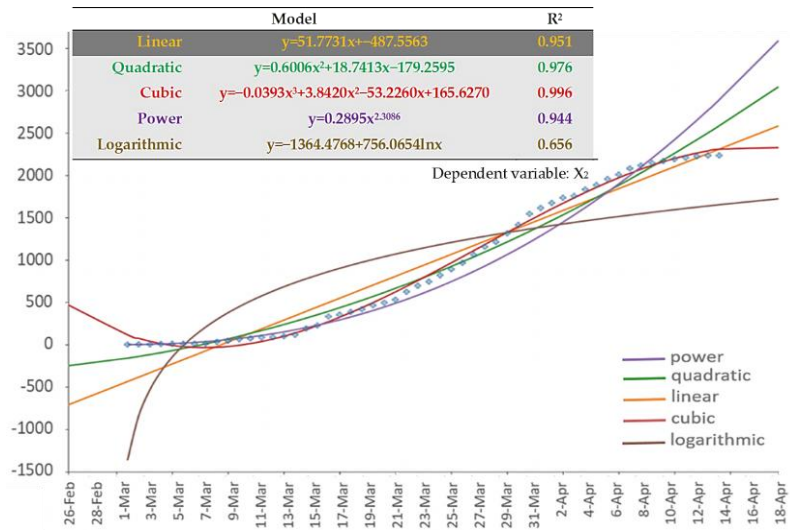

**Fig.6.** Available types of fitting curves applied to the cumulative COVID-19 infection curve (variable $X_2$) of Greece. Time-series data of the variable are shown in dots.

Next, Fig.7 shows the results of the regression analysis applied to the (cumulative) number of deaths (dependent variable: $X_4$). As it can be observed, similarly to variable $X_2$, the 3rd order polynomial (cubic) fitting curve best describes the data of the Greek COVID-19 cumulative deaths. The shape of this curve also implies that the number of cumulative infections tends to saturate.

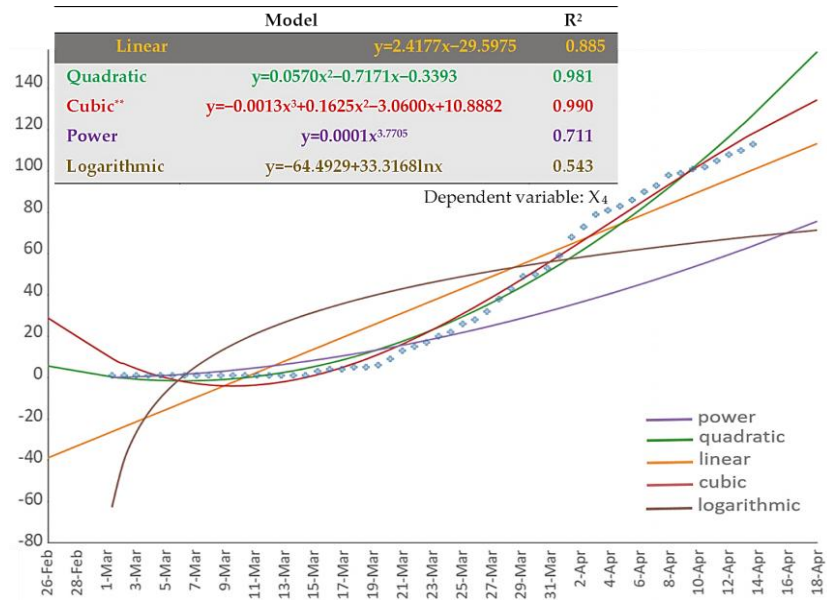

**Fig.7.** The available types of fitting curves applied to the cumulative COVID-19 death curve (variable $X_4$) of Greece. Time-series data of the variable are shown in dots.

Finally, Fig.8 shows the results of the regression analysis applied to the cumulative number of patients in ICU (dependent variable: cumulative $X_7$). Similarly to variables $X_2$ and $X_4$, the 3rd order polynomial (cubic) fitting curve best describes the data of (cumulative) variable $X_7$ and the shape of the curve implies that the number of cumulative ICU patients also tends to saturate.

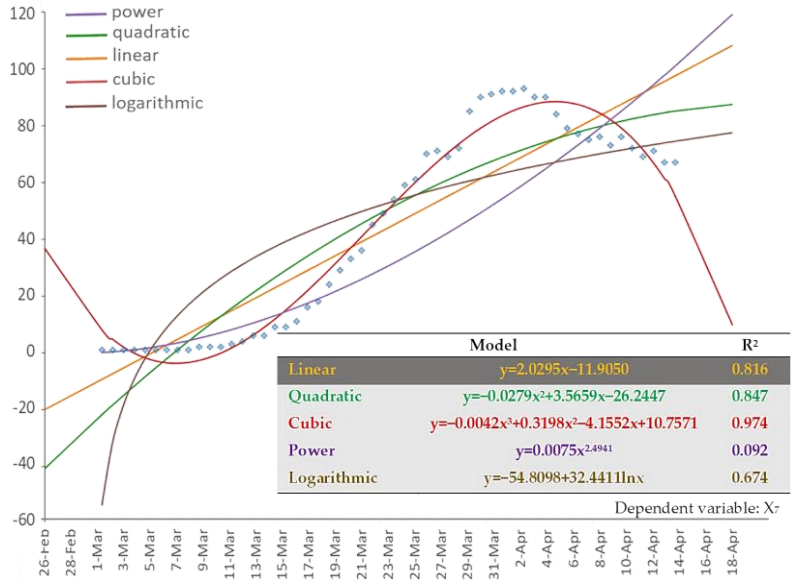

**Fig.8.** The available types of fitting curves applied to the cumulative COVID-19 ICU patients (variable: cumulative $X_7$) of Greece. Time-series data of the variable are shown in dots.

The regression analysis has shown that the best fitting to the cumulative expressions of the COVID-19 infection ($X_2$), death ($X_4$), and ICU patients ($X_7$) curves in Greece better fit to 3rd order polynomial (Cubic) than to linear, power, logarithmic, or 2nd order polynomial patterns. As it was previously observed, the cubic-shape of the fitting curves (which ends up to a convex area representing the recent past of the time-series) illustrates saturation trends of the COVID-19 evolution in Greece. To improve the accuracy and the determination ability of the fittings we apply at next a regression analysis based on the Regression Splines algorithm.

*4.2. Regression Splines*

A regression spline is a special piecewise polynomial function defined in parts and is widely used in interpolation problems requiring smoothing. In particular, for a given partition $a=t_0 < t_1 < t_2 < \ldots < t_{k-1} < t_n=b$ of the interval $[a,b]$, a spline is a multi-polynomial function $S(t)$ defined by the union of functions (De Boor et al., 1978):

$$S(t) = S_1([a,t_1]) \cup S_2([t_1,t_2]) \cup \ldots \cup S_{k-1}([t_{k-2},t_{k-1}]) \cup S_k([t_{k-1},b]) = \cup_{i=1}^{k} S_i([t_{i-1},t_i]) \quad (5),$$

where $k$ is the number of knots $\mathbf{t}=(t_0,t_1,t_2,\ldots,t_n)$ dividing the interval $[a,b]$ into $k$-1 convex subintervals. Each function $S_i(t)$, $i=1,\ldots,n$, is a polynomial of low (usually square) degree (sometimes can also be linear) that is fitter to the corresponding interval $[t_{i-1}, t_i]$, $i=1,\ldots,n$, so that the aggregate spline function to be continuous and smooth. The spline algorithm is preferable than this of simple regression in cases when the simple regression generates models of high degree (De Boor et al., 1978). This piecewise approach yields models of high determination by using low degree polynomial piece-functions. In terms of the bias-variance trade-off dilemma (Geman et al., 1992), stating that simple (i.e. of low degree) models have small variance and high bias whereas complex models have small bias and high variance, the spline algorithm can generate fittings of both low variance and low bias and thus it minimizes the expected loss expressed by the sum of square bias, variance, and noise.

The major modeling choices for applying splines is, first, the determination of the knot vector $\mathbf{t}=(t_0,t_1,t_2,\ldots,t_n)$ so that to obtain the smoothest and of best determination spline model, and, secondly, the selection of the polynomial degree so that the model to be smooth and continuous at the borders of the sub-intervals. Therefore, this highly effective (in terms of model determination) fitting method is very sensitive to the selection of the knot vector, which is usually being determined either uniformly, or arbitrarily, or intuitively, or based on the researchers' experience (De Boor et al., 1978; Geman et al., 1992). The more sophisticated knot-selection techniques existing in the literature (Li et al., 2004, 2005) build on heuristics to determine the knot vector generating the best fitting and smoothening of the spline model. Within this open debate of knot determination, this paper builds on the recent work of Tsiotas and Magafas (2020) and introduces a novel approach for the determination of the spline knot vector, based on complex network analysis. The proposed method is implemented on the COVID-19 infection curve of Greece and is compared with the previously best fitting models applied to variable $X_2$.

*4.3. Complex network analysis of time-series*

Transforming a time-series to a complex network is a modern approach that became recently popular, with the emergence of network science in various fields of research (Barabasi, 2016; Tsiotas and Charakopoulos, 2020). The most popular method to transform a complex network to a time-series is the visibility graph algorithm, which was proposed by Lacasa et al. (2008) and became dominant due to its intuitive conceptualization. In particular, the rationale of creating a time-series to a complex network (visibility graph) builds on considering the time-series as a path of successive mountains of different height (each representing the value of the time-series at the certain time). In this time-series-based landscape, an observer standing on a mountain can see (either forward or backwards) as far as no other mountain obstructs its visibility. In geometric terms, a visibility line can be drawn between two points (nodes) of the time-series whether no other intermediating node is higher than this pair of points and obstructs their visibility (Lacasa et al., 2008; Tsiotas and Magafas, 2020). Therefore, two time-series nodes can enjoy a connection in the associated visibility graph if they are visible through a visibility line (Lacasa et al., 2008). The visibility algorithm conceptualizes the time-series as a landscape and produces a visibility graph associated to this landscape (Tsiotas and Charakopoulos, 2020). The associated (to the time-series) visibility graph is a complex network where complex network analysis can be further applied (Tsiotas and Charakopoulos, 2020; Tsiotas and Magafas, 2020).

Within this context, by transforming the time-series of the COVID-19 infection curve to a visibility graph we can study the time-series as a complex network. This allows dividing the visibility graph of the COVID-19 infection into connective communities based on the modularity optimization algorithm of Blondel et al. (2008). This algorithm is heuristic and separates a complex network into communities which are dense within (i.e. links inside the communities are the highest possible) and sparse between (i.e. links inside the communities are the highest possible) the (Blondel et al., 2008; Fortunato, 2010; Tsiotas, 2019). Therefore, the most distant nodes within each community can define the knots for applying the spline algorithm. This complex-network-based definition (i.e. community detection based on modularity optimization) of the knot vector offers the missing conceptualization to the splines knots, defining them as borderline points of connectivity of the modularity-based communities. According to this approach, the visibility graph of the COVID-19 infection curve is divided to five modularity-based communities, which correspond to the periods Q1=[1,4] ∪ [9,19], Q2=[5,8], Q3=[20,26], Q4=[27,32], and Q5=[33,43], as it is shown in Fig.9, where positive integers in these intervals are elements of variable $X_1$.

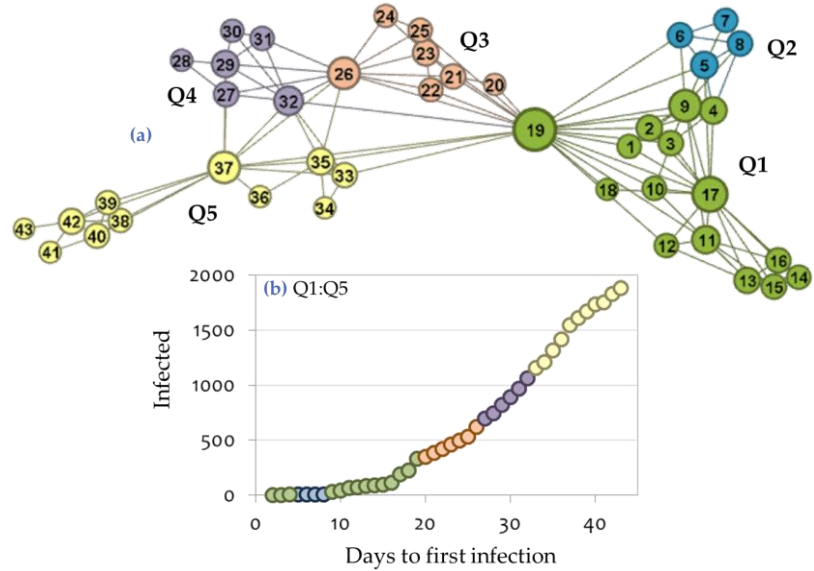

**Fig.9.** Community detection of the Greek COVID-19 infection visibility graph based on the modularity optimization algorithm of Blondel et al. (2008). Node size in the network is proportional to node degree.

Consequently, the spline knot vector can be defined by the knots $t$=(1,4,8,19,26,32,43) in the body of the time-series COVID-19 infection curve. This partition facilitates to apply the spline regression algorithm and to compare the determination ability of the spline model with the cubic regression models previously shown.

## 5. RESULTS AND DISCUSSION

After the complex-network-based determination of the spline knot vector, the spline regression algorithm is applied to the COVID-19 infection curve. The results are shown in Table 2, in comparison with the cubic fittings and with regression splines of randomly selected (3, 4 and 5) knots. As it can be observed, in all cases (i.e. for the dependent variables $X_2$, $X_4$, and $X_7$), the proposed complex-network spline models have better determination ability and lower error terms than both the cubic models resulted by the regression analysis and the randomly calibrated splines. In particular, improvements caused by the proposed method range between 0.00–0.20% for the multiple correlation coefficients ($R$), between 0.10–0.51% for the model determination ($R^2$), between 0.37–41.32% for the root mean square error (RMSE), and 0.25–34.19% for the relative absolute error (RAE). These improvements are considerable even in the cases of $R$ and $R^2$, given the already good fitting performance of the cubic and the randomly calibrated spline models.

**Table 2**
Comparison between the polynomial cubic and regression splines fitting curves

| Model | R | $R^2$ | RMSE* | RAE** |
|---|---|---|---|---|
| Dependent Variable: Infections ($X_2$) | | | | |
| Cubic | 0.998 | 0.996 | 2.229 | 4.182% |
| Regression Splines with 3 random knots | 0.999 | 0.998 | 1.805 | 3.798% |
| Regression Splines with 4 random knots | 0.999 | 0.998 | 1.621 | 3.277% |

| Model | R | R² | RMSE* | RAE** |
|---|---|---|---|---|
| Regression Splines with 5 random knots | 0.999 | 0.998 | 1.420 | 2.986% |
| **Complex-Network Regression Splines** | **1.000** | **1.000** | **1.308** | **2.752%** |
| Dependent Variable: Deaths ($X_4$) | | | | |
| Cubic | 0.995 | 0.990 | 2.423 | 4.981% |
| Regression Splines with 3 random knots | 0.995 | 0.990 | 2.584 | 5.013% |
| Regression Splines with 4 random knots | 0.995 | 0.990 | 2.423 | 4.798% |
| Regression Splines with 5 random knots | 0.995 | 0.990 | 2.410 | 4.732% |
| **Complex-Network Regression Splines** | **0.995** | **0.991** | **2.401** | **4.720%** |
| Dependent Variable: ICU patients ($X_7$) | | | | |
| Cubic | 0.987 | 0.974 | 6.300 | 12.659% |
| Regression Splines with 3 random knots | 0.987 | 0.974 | 6.287 | 12.648% |
| Regression Splines with 4 random knots | 0.988 | 0.976 | 6.186 | 12.114% |
| Regression Splines with 5 random knots | 0.989 | 0.979 | 6.204 | 12.007% |
| **Complex-Network Regression Splines** | **0.989** | **0.979** | **6.119** | **11.731%** |

*. Relative mean square error

**. Relative absolute error

According to these results, the proposed complex-network-based splines regression method outperforms in fitting determination both the cubic regression and the randomly-calibrated splines regression models, which are also models of high accuracy. In conceptual terms, this outperformance may be related with the immanent property of complex network analysis to model and manage problems of complexity and thus to provide better insights in the study complex systems, as in the case of the COVID-19 temporal spread. Despite the restriction in data availability, improvements (mainly in error terms) achieved by the proposed model are not negligible and highlight the direction of using hybrid or combined methodologies for dealing with cases of insufficient information. Moreover, the overall approach highlights the utility of multidisciplinary and synthetic modeling for dealing with complexity in epidemiology, which by default deals with complex socio-economic systems of humanity.

Given the small data availability, this improvement in the modeling determination is a promising advantage of the proposed method, which allows building on a quantitative consideration of the spline knot vector instead of an intuitive one. Besides, the utility and effectiveness of the proposed methodology should be evaluated in conjunction with the good performance of the spline results against multi-collinearity (Norusis, 2008; Walpole et al., 2012), which emerges in simple regression modeling due to the consideration of additional parameters.

Overall, the two major advantages of the proposed method concern the better stability and capability of forecasting, since the total behavior of the proposed model appears less noisy, while it reduces the errors of determination due to the modelers' choices. This conclusion can also be supported by the variance minimization of the expected error terms, which provides strong indications of the system's reliability.

## 6. CONCLUSIONS

Accurate forecasting is a major task in epidemiology that becomes very important today in the global emergence that the COVID-19 pandemic has caused. Due to the low availability of data, the worldwide conceptualization of the new pandemic is currently constraint and still emerging. Within the context of information lack, methods contributing to more accurate forecasting on early datasets are welcomed and pertinent for the ongoing fight against the disease. This paper proposed a novel method for modeling and forecasting in epidemiology based on complex network analysis and spline regression algorithm. Based on data of the COVID-19 temporal spread in Greece, the proposed method converted a time-series to an associated visibility graph and then it divided the graph into connected communities that defined the spline knot vector. This approach provided a complex-network definition of the spline knots, the definition of which is currently either intuitive or heuristic, and it assigned a conceptualization to the knots based on network connectivity.

Within this context, the proposed method was applied to different aspects of the COVID-19 temporal spread in Greece (the cumulative number of infections, deaths, and ICU patients) and found to outperform the regression cubic models, which had the highest determination amongst the available simple regression models. In methodological terms, the overall approach advances the spline regression algorithm, which is currently restricted to the not-well defined determination of knots, whereas, in practical implementation, the proposed methodology offers an active method for modeling and forecasting the pandemic, capable in removing the disconnected past data from the time-series structure. In terms of health policy, this paper provides a modeling and forecasting tool facilitating decision making and resource management in epidemiology, which can contribute in the ongoing fight against the pandemic of COVID-19.

**REFERENCES**

Ahmed, S. F., Quadeer, A. A., & McKay, M. R. (2020). Preliminary identification of potential vaccine targets for the COVID-19 coronavirus (SARS-CoV-2) based on SARS-CoV immunological studies. *Viruses*, *12*(3), 254.

Anastassopoulou, C., Russo, L., Tsakris, A., Siettos, C., (2020) "Data-based analysis, modeling and forecasting of the COVID-19 outbreak", *PLoS ONE*, 15(3), e0230405.

Bai, Y., Yao, L., Wei, T., Tian, F., Jin, D. Y., Chen, L., & Wang, M. (2020). Presumed asymptomatic carrier transmission of COVID-19. *Jama*, *323*(14), 1406-1407.

Barabási, A. L. (2016). *Network science*. Cambridge university press.

Blondel, V., Guillaume, J.-L., Lambiotte, R., Lefebvre, E., (2008) "Fast unfolding of communities in large networks", *Journal of Statistical Mechanics*, 10, P10008. https://doi.org/10.1088/1742-5468/2008/10/P10008.

De Boor, C., De Boor, C., Mathématicien, E. U., De Boor, C., & De Boor, C. (1978). *A practical guide to splines* (Vol. 27, p. 325). New York: springer-verlag.

Fang, Y., Zhang, H., Xie, J., Lin, M., Ying, L., Pang, P., & Ji, W. (2020). Sensitivity of chest CT for COVID-19: comparison to RT-PCR. *Radiology*, 200432.

Fong, S.j., Li, G., Dey, N., Crespo, R. G., Herrera-Viedma, E., (2020) "Finding an Accurate Early Forecasting Model from Small Dataset: A Case of 2019-nCoV Novel Coronavirus Outbreak", *International Journal of Interactive Multimedia and Artificial Intelligence*, 6(1), pp.132-40.

Fortunato, S., (2010) "Community detection in graphs", Physics Reports, 486, pp.75–174. https://doi.org/10.1016/j.physrep.2009.11.002.

Geman, S., Bienenstock, E., & Doursat, R. (1992). Neural networks and the bias/variance dilemma. *Neural computation*, *4*(1), 1-58.

Giuliani, D., Dickson, M.-M., Espa, G., and Santi, F., (2020) "Modelling and Predicting the Spatio - Temporal Spread of Coronavirus Disease 2019 (COVID -19) in Italy", (10.2139/ssrn.3559569).

Heymann, D. L., & Shindo, N. (2020). COVID-19: what is next for public health?. *The Lancet*, *395*(10224), 542-545.

Lacasa, L., Luque, B., Ballesteros, F., Luque, J., Nuno, J.C., (2008) "From time-series to complex networks: The visibility graph", *Proceedings of the National Academy of Sciences*, 105(13), pp.4972–4975.

Lescure, F. X., Bouadma, L., Nguyen, D., Parisey, M., Wicky, P. H., Behillil, S., ... & Enouf, V. (2020). Clinical and virological data of the first cases of COVID-19 in Europe: a case series. *The Lancet Infectious Diseases*.

Li, W., Xu, S., Zhao, G., & Goh, L. P. (2004). A heuristic knot placement algorithm for B-spline curve approximation. *Computer-Aided Design and Applications*, *1*(1-4), 727-732.

Li, W., Xu, S., Zhao, G., & Goh, L. P. (2005). Adaptive knot placement in B-spline curve approximation. *Computer-Aided Design*, *37*(8), 791-797.

Liu, W., Zhang, Q., Chen, J., Xiang, R., Song, H., Shu, S., ... & Wu, P. (2020). Detection of Covid-19 in children in early January 2020 in Wuhan, China. *New England Journal of Medicine*, *382*(14), 1370-1371.

Livingston, E., Bucher, K., (2020) "Coronavirus disease 2019 (COVID-19) in Italy", Jama.

Lurie, N., Saville, M., Hatchett, R., & Halton, J. (2020). Developing Covid-19 vaccines at pandemic speed. *New England Journal of Medicine*.

Mahase, E., (2020) "COVID-19: UK starts social distancing after new model points to 260 000 potential deaths", BMJ, 368:m1089.

Ministry of Health of Greece – MOHG (2020), "Press Releases", available at the URL: https://www.moh.gov.gr/articles/ministry/grafeio-typoy/press-releases [accessed: 30/4/20].

National Public Health Organization of Greece – NPHOG (2020) "New coronavirus Covid-19 - Instructions", available at the URL: https://eody.gov.gr/neos-koronaios-covid-19 [accessed: 26/4/2020].

Norusis, M., (2008) SPSS 16.0 advanced statistical procedures companion, Prentice Hall Press.

Our World in Data, (2020) "Total confirmed COVID-19 deaths per million people", available at the URL: https://ourworldindata.org/grapher/total-covid-deaths-per-million?year=2020-04-26 [accessed: 26/4/2020].

Petropoulos, F., Makridakis, S., (2020) "Forecasting the novel coronavirus COVID-19", *PLoS ONE*, 15(3), e0231236.

Qiu, H., Wu, J., Hong, L., Luo, Y., Song, Q., & Chen, D. (2020). Clinical and epidemiological features of 36 children with coronavirus disease 2019 (COVID-19) in Zhejiang, China: an observational cohort study. *The Lancet Infectious Diseases*.

Remuzzi, A., Remuzzi, G., (2020) "COVID-19 and Italy: what next?", The Lancet.

Roser, M., Ritchie, H., (2020) "Coronavirus Disease (COVID-19)" available at the URL: https://ourworldindata.org/coronavirus-data [accessed: 10/4/20]

Sarkodie, S.A., Owusu, P.A., (2020) "Investigating the cases of novel coronavirus disease (COVID-19) in China using dynamic statistical techniques", *Heliyon*, 6(4), e03747.

Shi, H., Han, X., Jiang, N., Cao, Y., Alwalid, O., Gu, J., ... & Zheng, C. (2020). Radiological findings from 81 patients with COVID-19 pneumonia in Wuhan, China: a descriptive study. *The Lancet Infectious Diseases*.

Tan, L., Wang, Q., Zhang, D., Ding, J., Huang, Q., Tang, Y. Q., ... & Miao, H. (2020). Lymphopenia predicts disease severity of COVID-19: a descriptive and predictive study. *Signal transduction and targeted therapy*, 5(1), 1-3.

Tsiotas, D., Charakopoulos, A., (2020) "VisExpA: Visibility expansion algorithm in the topology of complex networks", *Software X*, 11, 100379.

Tsiotas, D., Magafas, L., (2020) "The effect of anti-COVID-19 policies to the evolution of the disease: A complex network analysis to the successful case of Greece", arXiv.2004.06536.

Walpole, R. E., Myers, R. H., Myers, S. L., Ye, K., (2012) *Probability & Statistics for Engineers & Scientists*, ninth ed., New York, Prentice Hall Publications.

Worldometers, (2020) "COVID-19 coronavirus pandemic", available at the URL: www.worldometers.info/coronavirus [accessed: 20/4/2020].

Xu, B., Gutierrez, B., Mekaru, S. et al. (2020) "Epidemiological data from the COVID-19 outbreak, real-time case information", *Sci Data*, 7(106) (doi:10.1038/s41597-020-0448-0).

Xu, B., Gutierrez, B., Mekaru, S., Sewalk, K., Goodwin, L., Loskill, A., ... , Zarebski, A. E., (2020b) "Epidemiological data from the COVID-19 outbreak, real-time case information", *Scientific Data*, 7(1), pp.1-6.

Xu, Z., Shi, L., Wang, Y., Zhang, J., Huang, L., Zhang, C., ... & Tai, Y. (2020a). Pathological findings of COVID-19 associated with acute respiratory distress syndrome. *The Lancet respiratory medicine*, 8(4), 420-422.

Ye, Z., Zhang, Y., Wang, Y., Huang, Z., & Song, B. (2020). Chest CT manifestations of new coronavirus disease 2019 (COVID-19): a pictorial review. *European Radiology*, 1-9.

## APPENDIX A

**Table A1**

The COVID-19 variables participating in the analysis[a]

| Date | Day ($X_1$) | Infections ($X_2$) | Daily Infections ($X_3$) | Deaths ($X_4$) | Daily Deaths ($X_5$) | Daily Recovered ($X_6$) | ICU ($X_7$) | Tests ($X_8$) |
|---|---|---|---|---|---|---|---|---|
| 26-Feb-20 | 1 | 1 | 1 | 0 | 0 | 0 | 0 | 0 |
| 27-Feb-20 | 2 | 3 | 2 | 0 | 0 | 0 | 0 | 0 |
| 28-Feb-20 | 3 | 4 | 1 | 0 | 0 | 0 | 0 | 0 |
| 29-Feb-20 | 4 | 7 | 3 | 0 | 0 | 0 | 0 | 0 |
| 1-Mar-20 | 5 | 9 | 0 | 0 | 0 | 0 | 0 | 0 |
| 2-Mar-20 | 6 | 9 | 0 | 0 | 0 | 0 | 0 | 0 |
| 3-Mar-20 | 7 | 9 | 0 | 0 | 0 | 0 | 0 | 270 |
| 4-Mar-20 | 8 | 11 | 2 | 0 | 0 | 0 | 1 | 0 |
| 5-Mar-20 | 9 | 31 | 22 | 0 | 0 | 0 | 0 | 0 |
| 6-Mar-20 | 10 | 46 | 14 | 0 | 0 | 0 | 1 | 0 |

| Date | Day (X₁) | Infections (X₂) | Daily Infections (X₃) | Deaths (X₄) | Daily Deaths (X₅) | Daily Recovered (X₆) | ICU (X₇) | Tests (X₈) |
|---|---|---|---|---|---|---|---|---|
| 7-Mar-20 | 11 | 66 | 21 | 0 | 0 | 0 | 0 | 0 |
| 8-Mar-20 | 12 | 73 | 7 | 0 | 0 | 0 | 0 | 0 |
| 9-Mar-20 | 13 | 84 | 11 | 0 | 0 | 0 | 1 | 0 |
| 10-Mar-20 | 14 | 89 | 5 | 0 | 0 | 0 | 1 | 0 |
| 11-Mar-20 | 15 | 99 | 10 | 0 | 0 | 0 | 2 | 0 |
| 12-Mar-20 | 16 | 117 | 18 | 0 | 1 | 2 | 0 | 1910 |
| 13-Mar-20 | 17 | 190 | 73 | 1 | 0 | 0 | 3 | 520 |
| 14-Mar-20 | 18 | 228 | 38 | 1 | 2 | 6 | 0 | 700 |
| 15-Mar-20 | 19 | 331 | 103 | 3 | 1 | 2 | 2 | 0 |
| 16-Mar-20 | 20 | 352 | 21 | 4 | 0 | 0 | 5 | 920 |
| 17-Mar-20 | 21 | 387 | 35 | 4 | 1 | 4 | 2 | 580 |
| 18-Mar-20 | 22 | 418 | 31 | 5 | 0 | 0 | 6 | 1100 |
| 19-Mar-20 | 23 | 464 | 46 | 5 | 1 | 5 | 5 | 300 |
| 20-Mar-20 | 24 | 495 | 31 | 6 | 3 | 0 | 4 | 872 |
| 21-Mar-20 | 25 | 530 | 35 | 9 | 4 | 0 | 3 | 658 |
| 22-Mar-20 | 26 | 624 | 94 | 13 | 2 | 0 | 9 | 176 |
| 23-Mar-20 | 27 | 695 | 71 | 15 | 2 | 10 | 4 | 638 |
| 24-Mar-20 | 28 | 743 | 48 | 17 | 3 | 3 | 5 | 427 |
| 25-Mar-20 | 29 | 821 | 78 | 20 | 2 | 4 | 5 | 1424 |
| 26-Mar-20 | 30 | 892 | 71 | 22 | 4 | 6 | 2 | 0 |
| 27-Mar-20 | 31 | 966 | 74 | 26 | 2 | 10 | 9 | 2982 |
| 28-Mar-20 | 32 | 1061 | 95 | 28 | 4 | 0 | 1 | 886 |
| 29-Mar-20 | 33 | 1156 | 95 | 32 | 6 | 0 | -2 | 788 |
| 30-Mar-20 | 34 | 1212 | 56 | 38 | 5 | 0 | 3 | 810 |
| 31-Mar-20 | 35 | 1314 | 102 | 43 | 6 | 0 | 13 | 771 |
| 1-Apr-20 | 36 | 1415 | 101 | 49 | 1 | 0 | 5 | 618 |
| 2-Apr-20 | 37 | 1544 | 129 | 50 | 3 | 9 | 1 | 1494 |
| 3-Apr-20 | 38 | 1613 | 69 | 53 | 6 | 17 | 1 | 3593 |
| 4-Apr-20 | 39 | 1673 | 60 | 59 | 9 | 0 | 0 | 896 |
| 5-Apr-20 | 40 | 1735 | 62 | 68 | 5 | 0 | 1 | 2120 |
| 6-Apr-20 | 41 | 1755 | 20 | 73 | 6 | 191 | -3 | 740 |
| 7-Apr-20 | 42 | 1832 | 77 | 79 | 2 | 0 | 0 | 2391 |
| 8-Apr-20 | 43 | 1884 | 52 | 81 | 2 | 0 | -6 | 3944 |
| 9-Apr-20 | 44 | 1955 | 71 | 83 | 3 | 0 | -5 | 1106 |
| 10-Apr-20 | 45 | 2009 | 56 | 86 | 4 | 0 | -2 | 1798 |
| 11-Apr-20 | 46 | 2081 | 70 | 90 | 3 | 0 | -2 | 1912 |
| 12-Apr-20 | 47 | 2114 | 33 | 93 | 5 | 0 | 1 | 4917 |
| 13-Apr-20 | 48 | 2145 | 31 | 98 | 1 | 0 | -3 | 1156 |
| 14-Apr-20 | 49 | 2170 | 25 | 99 | 2 | 0 | 3 | 5381 |
| 15-Apr-20 | 50 | 2192 | 22 | 101 | 1 | 0 | -4 | 1973 |
| 16-Apr-20 | 51 | 2207 | 15 | 102 | 3 | 0 | -3 | 0 |
| 17-Apr-20 | 52 | 2224 | 17 | 105 | 3 | 0 | 2 | 0 |
| 18-Apr-20 | 53 | 2235 | 11 | 108 | 2 | 0 | -4 | 2519 |
| 19-Apr-20 | 54 | 2235 | 0 | 110 | 3 | 0 | -4 | 0 |
| **Sum** | | **2235** | **2235** | **113** | **113** | **269** | **69** | **53290** |

*. All data were daily extracted from National Public Health Organization of Greece (2020)